# Phonon Directionality Impacts Electron-Phonon Coupling and Polarization of the Band-Edge Emission in Two-Dimensional Metal Halide Perovskites

*Roman Krahne[1], Alexander Schleusener[1], Mehrdad Faraji[1,2], Lin-Han Li[3,4], Miao-Ling Lin[3,4], and Ping-Heng Tan[3,4].*

Affiliations

[1] Optoelectronics Research Line, Istituto Italiano di Tecnologia (IIT), Via Morego 30, 16163 Genoa, Italy

[2] Dipartimento di Chimica e Chimica Industriale, Università degli Studi di Genova, Via Dodecaneso, 31, 16146 Genova, Italy

[3] State Key Laboratory of Superlattices and Microstructures, Institute of Semiconductors, Chinese Academy of Sciences, 100083 Beijing, China

[4] Center of Materials Science and Optoelectronics Engineering & CAS Center of Excellence in Topological Quantum Computation, University of Chinese Academy of Sciences, 100049 Beijing, China.

Two-dimensional metal-halide perovskites are highly versatile for light-driven applications due to their exceptional variety in material composition, which can be exploited for tunability of mechanical and optoelectronic properties. The band edge emission is defined by structure and composition of both organic and inorganic layers, and electron-phonon coupling plays a crucial role in the recombination dynamics. However, the nature of the electron-phonon coupling and which kind of phonons are involved is still under debate. Here we investigate the emission, reflectance and phonon response from single two-dimensional lead-iodide microcrystals with angle-resolved polarized spectroscopy. We find an intricate dependence of the emission polarization with the vibrational directionality in the materials, which reveals that several bands of the low-frequency phonons with non-orthogonal directionality contribute to the band edge emission. Such complex electron-phonon coupling requires adequate models to predict the thermal broadening of the emission and provides opportunities to design its polarization properties.

Two-dimensional layered organic-inorganic perovskites alternate sub-nanometer thin semiconductor layers consisting of a single plane of inorganic metal halide octahedra with dielectric layers formed by the organic cations.[1] This leads to strong confinement of the electronic carriers, which renders such systems as natural quantum wells, with binding energies for electron-hole pairs (excitons) of the order of several hundred meV.[2-5] These properties are highly appealing for light emission and manipulation, in particular, because the emission can be tailored by the choice of the halide, and to some extent by the composition of the organic layers that influence the structural and electrical properties of the material.[6-8] Here the static distortions induced by the binding of the organic cations to the octahedra layer have a significant impact on the crystal structure and electron-phonon coupling.[6] Furthermore the organic layer defines the dielectric confinement of the charge carriers in the semiconductor metal halide lattice. Several recent works investigated the confinement effects on the exciton fine structure,[9-13] and provided a detailed understanding with insights on crystal symmetry, polar distortions, and other structural property effects.[11, 12, 14-18] However, the influence of the electron-phonon coupling in the recombination dynamics is much less investigated, and which mechanisms and phonon bands contribute to the band edge emission is under debate. For example, a series of equally spaced peaks in the emission spectra has been attributed to a vibronic progression due to coupling to phonons of the organic cations[19-21], while other works observed emission dynamics that were related to polaron formation.[22-26]

In this work, we correlate the band-edge exciton emission of single two-dimensional lead-iodide perovskite crystals with their vibrational properties, where we measure the phonon response by Raman spectroscopy under resonant (above band gap) and non-resonant (below band gap) excitation. We record angle-resolved polarized photoluminescence (PL), reflectance and Raman spectra from the same region of single, regularly shaped 2D microcrystals, where we resolve emission peaks with a width of less than 0.5 meV, and measure the low-frequency Raman response down to 8 cm$^{-1}$ (equivalent to 1 meV). This allows to directly correlate the in-plane polarization of the different exciton emission bands with the directionality of the phonon modes of the inorganic octahedra lattice. We find a distinct polarization of the different emission peaks that is strongly connected to the directionality of the low-frequency phonon bands. The phonon spectrum and directionality of the vibrational modes depends on the type of organic cation in the 2D crystal. In particular, in some materials dominant phonon modes occur along the principal directions of the octahedra lattice and along their diagonals, which enables contributions of combinations of phonons with different directionality in the electron-phonon coupling.

RESULTS & DISCUSSION

**Figure 1 (A-B)** illustrates the main features of the optical setup that enables angle-resolved PL and Raman spectroscopy of single micron-sized crystals. The coaxial beam path of the different laser lines allows to excite the samples with different wavelengths (457 nm for above, and 633 nm for below band gap excitation), and the switching of the grating in the spectrometer enables to record PL and Raman spectra from the same excitation spot. For reflectance measurements, a white light source is used instead of the laser. In the experiments, we first performed the PL and reflectance measurements that required much shorter integration time (fractions of a second) and less excitation power, and then the Raman experiments in which the signal was integrated over several tens of seconds and, if needed, measured with higher excitation power. In all cases, we confirmed that the beam exposure did not lead to degradation or damage of the sample by optical imaging of the microcrystals and inspection of their optical spectra.

As metal halide perovskite samples, we chose two-dimensional lead-iodide crystals in the Ruddlesden-Popper phase with butylammonium (BA), undecylammonium (UDA), and phenylethylammonium (PEA) as organic cations. $BA_2PbI_4$ is the most studied 2DLP system with an aliphatic chain cation, and $UDA_2PbI_4$ provides insight into the impact of the aliphatic chain length on the optoelectronic properties. $PEA_2PbI_4$ is the archetype of a 2D perovskite with a phenyl ring in the organic portion that leads to different properties of the organic layer, for example in its stiffness. Furthermore, the above materials have different symmetries in their low-temperature phase (see Supporting Information (SI) Section 1 for a discussion of their structural properties).

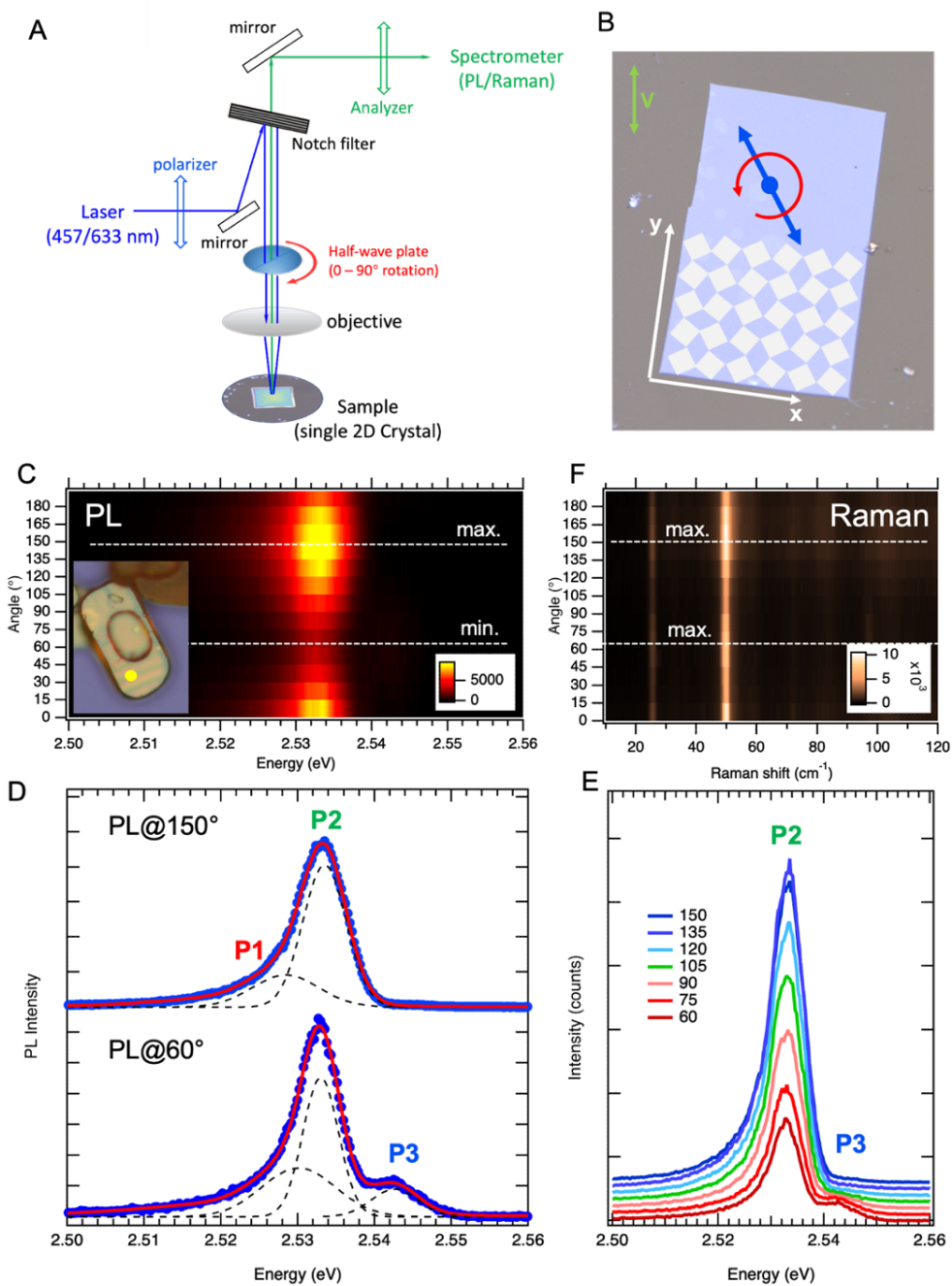

**Figure 1.** Directional photoluminescence (PL) and Raman spectra from single 2D perovskite flakes. A) Scheme illustrating the setup for the angle-resolved PL and Raman spectroscopy

experiments. The half-wave plate in the combined excitation and collection path allows to rotate the angle of the linear polarization of the light with respect to the orientation of the exfoliated single perovskite crystal under investigation. PL and Raman spectroscopy measurements were recorded from the same micron-size spot of the 2D perovskite flakes. B) Optical microscope image of a single exfoliated 2D perovskite flake overlayed with an illustration of its octahedra lattice. The blue arrow illustrates the linear polarization impinging on the sample, which is defined by the angle of the half wave plate that can be rotated with respect to the orientation of the flake (red circular arrow) and the vertical axis of the setup (green arrow) that corresponds to the analyzer direction. C) Color map of the angle-dependent PL intensity with laser excitation at 457 nm (2.71 eV). The inset shows an optical image of the measured flake, where the yellow dot marks the location of the excitation spot, and the green arrow indicates the direction of the (fixed) vertical analyzer of the optical setup. D) PL spectra recorded at angles of 60° and 150° (white dashed lines in A) together with Gaussian fitting of the spectra. E) Waterfall plot of the spectra recorded from 60° (red) to 150° (blue), which show the increase of P2 and decrease of P1 with increasing angle. F) Angle-resolved resonant Raman spectra displayed as a color map, recorded with 457 nm laser excitation from the same spot as the PL. See Figures S3 and S4 in Section 2 of the SI for non-resonant Raman spectroscopy data and reflectance measurements.

The angle-resolved PL spectra measured from a single $BA_2PbI_4$ flake are shown in Figure 1C. We clearly observe a maximum in the overall emission intensity at an angle of around 150° and a minimum at around 60°. These angles correspond to the orientation of the flake, and therefore we can associate the main emission intensity roughly to be along one of the major axes of the octahedra lattice. Gaussian fitting of the emission spectra reveals the contributions of two bands to the main peak, as depicted in Figure 1D, where the spectra recorded at 60° and 150° are displayed. The emission at 150° is characterized by a Gaussian peak (P2) at 2.534 eV with a width of 4 meV (we take as Gaussian peak width twice the standard deviation) that has a broader shoulder fitted by Gaussian (6 meV width) on its low energy side that we label P1. The spectrum at 60° is consists of three Gaussian peaks, two with similar narrow width of around 4 meV, and a low energy shoulder with 7 meV width. We associate the peaks with ca 4 meV width to the confined band edge excitons, since their polarization behavior in PL and reflectance (see Figure S4) agree well with reports in literature. [9, 18, 27, 28] We focus on the peaks labeled P2 and P3 that show an orthogonal behavior in intensity with respect to the polarization angle. Fig. 2C shows that P3 decreases from 60° to 150°, while P2 increases. Therefore, we assign P2 and P3 to the confined excitons along the orthogonal in-plane directions of the major axes of the 2D single crystal. The resonant Raman spectra recorded from the same region are depicted in Figure 1F. Dominant modes are found at 25 $cm^{-1}$ and 49 $cm^{-1}$, which have their intensity maxima at the same angles as the PL maxima of peaks P2 and P3 (indicated by the white dashed lines in Figure 1C,F). We therefore conclude that these low-frequency phonons, which correspond to octahedral twists (25 $cm^{-1}$) and Pb-I bond bending and stretching (49 $cm^{-1}$), [22, 29-33] are fundamentally involved in the emission dynamics. To summarize, the regular $BA_2PbI_4$ flake manifests two emission bands, P2 and P3, respectively, that have their intensity maxima at orthogonal angles, and which can be associated to the in-plane confined exciton modes along the major axis of the octahedra lattice. Intensity maxima of the low-frequency phonon modes occur at the same angles as the emission intensity maxima, pointing to a strong impact of electron-phonon coupling with these low-frequency vibrational modes in the recombination dynamics, in agreement with literature. [26]

We now turn to $UDA_2PbI_4$ that features an organic cation with a longer aliphatic chain (11 carbons), which leads to a thicker organic layer and stronger dielectric confinement. Accordingly, the emission of UDA$_2$PbI$_4$ is blue-shifted with respect to BA$_2$PbI$_4$, as shown in Figure S2 in the SI. Furthermore, from literature we expect the low temperature crystal phase of UDA$_2$PbI$_4$ to be monoclinic, with a stronger out-of-plane octahedra tilting. [34]

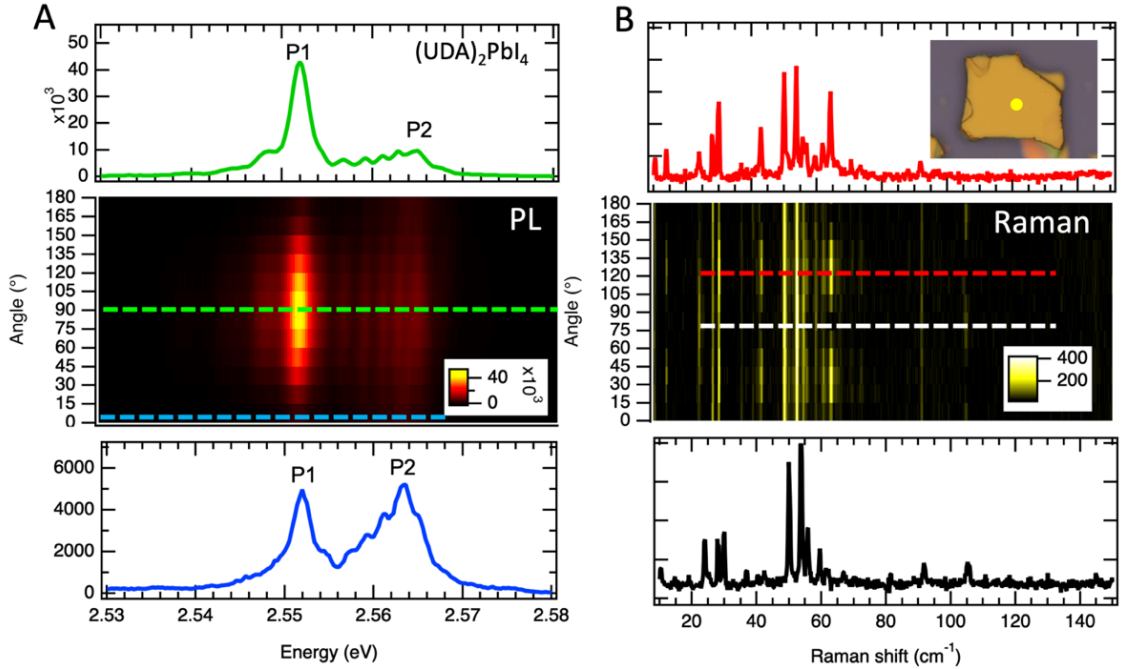

**Figure 2.** Angle-resolved polarized PL and Raman experiments from a single UDA$_2$PbI$_4$ microcrystal recorded at T=4K. A) Color plot of the angle-dependent PL spectra, the dashed green and blue lines indicate the angle of the maximum (green) and minimum (blue) PL intensity, and the respective spectra are displayed on the top and bottom of the color plot. Excitation wavelength was 457 nm. B) Color plot of the angle-dependent Raman spectra excited non-resonantly with 633 nm wavelength. The red (white) dashed lines indicate angles separated by 45° at which several modes have their maximum intensity. Spectra recorded at 120° (red dashed line) and 75° (white dashed line are shown on the top and bottom of the color plot, respectively. The inset shows an optical microscope image of the investigated microcrystal where the measurement spot is indicated by the yellow dot. Resonant Raman spectra are reported in Figure S5 in the SI.

**Figure 2A** displays the angle-resolved PL of a single UDA2PbI$_4$ flake. In contrast to BA$_2$PbI$_4$, all emission peaks in the UDA$_2$PbI$_4$ spectrum have their intensity maximum at the same angle, at around 90°, and we observe a more complex emission spectrum that contains a large number of peaks. Focusing on the two main bands that we label as P1 and P2, we can distinguish differences in the strength of the intensity increase with changing angle: P1 is slightly weaker than P2 at 0°, which is reversed at 90° by the strong angle-dependent increase in intensity of P1 (displayed in detail in Figure S6). The polarization dependence of the emission peaks with a single angular maximum goes along with phonon modes under resonant excitation that manifest no significant angle-dependent intensity variation, see Figure S7. However, we find a very interesting angle-dependent behavior in the Raman spectra recorded non-resonantly, thus with excitation below the band gap at 633nm, as shown in Figure 2B. Here we observe a very rich phonon spectrum with a large number of below 120 cm$^{-1}$ (15 meV), where a set of prominent modes has their maxima at 30° and 120°, another set of strong modes has its maximum at 75°, and again another set has two maxima at 75° and 165°. These findings are in very good agreement with our data from UDA$_2$PbBr$_4$ microcrystals,[33] and with the regular shape of the microcrystal displayed in the inset of Figure 2B we can assign the modes that peak at 75° vibrations along the [110] direction of the crystal lattice, those at 165° to the orthogonal [-110] direction, but most interestingly, the sets at 30° and 120° to diagonal lattice directions, that is along [100] and [010]. We note that for some microcrystals such diagonal lattice direction can also be distinguished by well-defined 45° angles of their edges (see Figure S8 for an example). One interesting point is that phonon modes that differ by 45° angles have non-zero vector products and therefore coupled modes can participate in the optical recombination processes. This enables a new range electron-phonon coupling processes that go beyond the typically considered Froehlich coupling involving only one longitudinal optical phonon mode.[35] Furthermore, the coupling of phonon modes along different in-plane lattice directions can lead to a single angle-dependent emission intensity maximum, in agreement with our data in Figure 2A.

The angle-dependent polarized reflectance and PL spectra of a different UDA$_2$PbI$_4$ microcrystal are shown in Figure 3 (see Figure S9 for the full angular data set). Here the orientation of the microcrystal with the vertical axis is around 45°, and accordingly, we observe the minima and maxima in PL intensity at 45° and 150° angles, respectively. In reflectance, we find a series of minima indicated by the dashed black lines that depend on the in-plane polarization angle. The minimum marked by the magenta dashed line (at 2.576 eV) does not change with angle, and therefore we attribute it to the exciton state confined in vertical ($z$) direction. Since the fine structure splitting is expected to be only a few meV, [18] we attribute the small dip at 2.573 eV to the confined exciton state in $y$-direction, which is corroborated by its decrease in depth with decreasing angle, while at the same time a minimum at 2.571 eV develops, which consequently can be assigned to the $x$-confined exciton state.[27, 28] The other three, more pronounced, dips in reflectance could be related to exciton polaritons that originate from the open cavity formed by the microcrystal itself, as reported in ref. [36]. In this scenario, the redshift with decreasing angle would result from the difference of the confined exciton energies in the two orthogonal directions.

The strong PL band centered at 2.545 eV does not have a counterpart in reflectance, and does not shift in energy with the in-plane polarization angle. Thus we assign it to a phonon sideband of the band edge emission. Its spectral shape resembles that of the (polariton) peak at 2.558 eV, and thus it is reasonable to assume that it is a phonon replica of this emission band, which would result in an energy spacing of ca 15 meV (120 cm$^{-1}$). We do not observe any phonon bands at such energy in our Raman experiments (Figure 2B), however, this energy could be explained by the sum of the dominant phonon energies (around 60 cm$^{-1}$) that occur in directions shifted by 45° angles. Therefore, such large energy shift supports our reasoning that combinations (vector products) of the directional phonons are involved in the electron-phonon coupling.

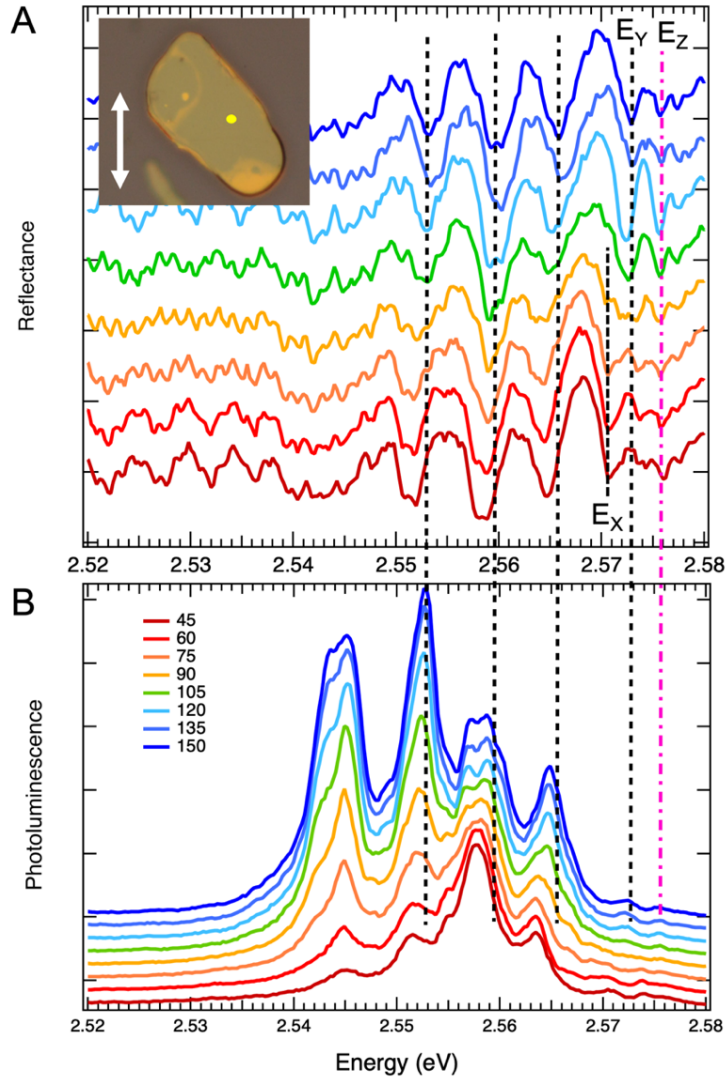

**Figure 3.** Angle-resolved polarized reflectance (A) and PL (B) spectra recorded from a single UDA$_2$PbI$_4$ microcrystal deposited on a quartz substrate. The color coding for the angles from 45°

to 150° is detailed in the legend of (B) and applies to both panels. The temperature 7K. The dashed lines indicate the minima in reflectance and are a guide to the eye to correlate the PL peaks to the reflectance dips. $E_X$, $E_Y$, and $E_Z$ denote the confined exciton levels.

We also investigated the emission, reflectance and vibrational properties of single $PEA_2PbI_4$ flakes and report the PL and Raman spectra in Section 4 in the SI. Overall, the behavior of $PEA_2PbI_4$ is similar to that of $UDA_2PbI_4$, featuring intensity maxima of the emission peaks at a single polarization angle, and phonon bands with a strong angular intensity variation under non-resonant excitation in the Raman spectra that comprise dominant modes in the [110] and [100] crystal lattice directions (Figures S10-S12). The main emission band shifts in energy with the in-plane polarization angle, in agreement with literature,[9] and we can identify the confined exciton modes in $x$ and $y$ directions with reflectance measurements (see Figure S13) as reported in ref. [28] Interestingly, for some microcrystals, we observe two additional strong sharp peaks in the emission spectrum, and their energy spacing of 9.4 meV corresponds very well to the energy of the out-of-plane phonon mode, see Figure S10. Therefore, we conclude that these sharp peaks originate from a vibronic progression related to this out-of-plane phonon mode that is favored by the stiffer organic layer in $PEA_2PbI_4$, in agreement with literature. [20, 21] This interpretation is corroborated by the weak angular intensity dependence of both these sharp emission peaks and of the out-of-plane phonon mode.

**Figure 4** reports temperature dependent PL experiments in the range from 4-80K that reveal how the band edge emission observed at 4K evolves from a single broad peak at 80K for all samples. The exciton-fine structure emerges in the temperature range below 80K, which underpins the key role of the low-frequency phonons with energy below 10 meV in the emission dynamics. With decreasing temperature, we observe a red shift of the main emission band for BA$_2$PbI$_4$ (Figure 4A) and PEA$_2$PbI$_4$ (Figure 4C,D), in agreement with literature, [37, 38] while for UDA$_2$PbI$_4$ (Figure 4B) the broad peak observed in the range from 60 - 80 K splits into a set of sharper peaks below 40K. For the systems build with organic cations with an aliphatic chain (BA$_2$PbI$_4$ and UDA$_2$PbI$_4$), the emission intensity increases significantly with decreasing temperature. This is different for PEA$_2$PbI$_4$, where the emission intensity has a maximum at around 50K, a local minimum at 20K, and then again relatively strong emission intensity at 4K. Such behavior can be explained by the involvement of phonons with different energies in the emission process, namely the out-of-plane phonon at 9.4 meV (77 cm$^{-1}$), and the low-frequency in-plane phonons in the range from 2.5 -8 meV (20-70 cm$^{-1}$). The weak dependence of the emission intensity at temperatures below 30K for PEA$_2$PbI$_4$ indicates that the impact of the in-plane phonons to the thermal coupling of the exciton band edge states is much less as compared to BA$_2$PbI$_4$ and UDA$_2$PbI$_4$. Accordingly, we do not find a low-frequency signal in the resonant Raman spectra recorded from the PEA$_2$PbI$_4$ flake, see Figure S10.

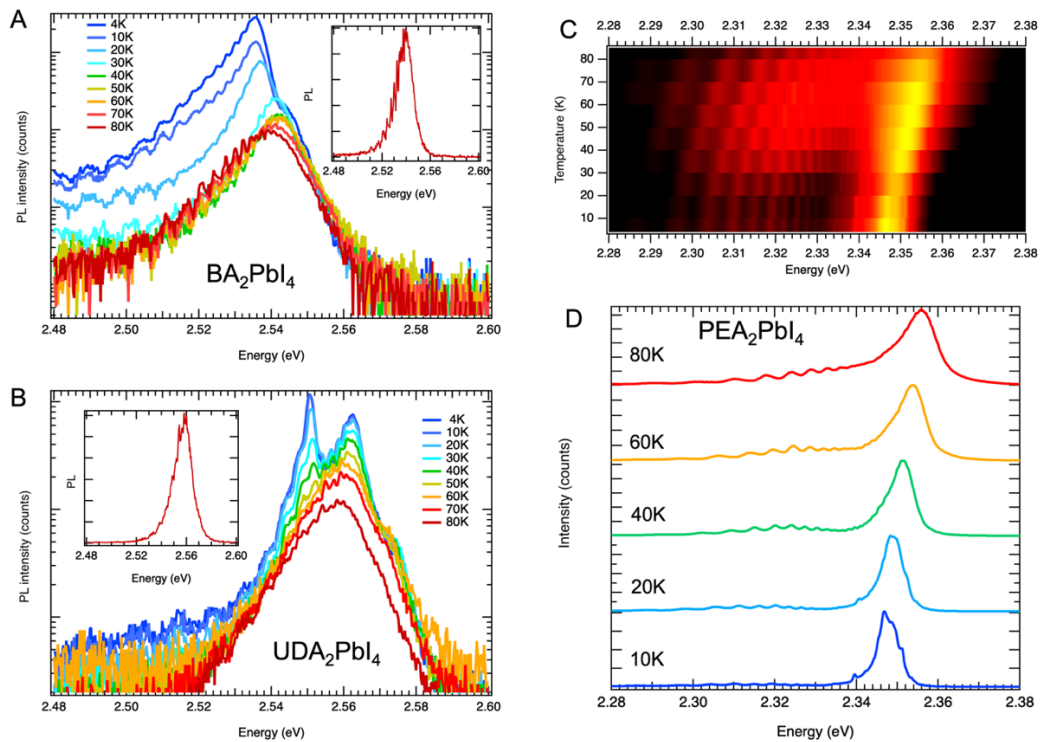

**Figure 4.** Temperature dependence of the band-edge emission of single lead-iodide microcrystals. A-B) Semi-log plot of the emission intensity of $BA_2PbI_4$ (A) and $UDA_2PbI_4$ (B). The insets show the emission at 80K on a linear scale. C-D) Color-plot and emission spectra at different temperatures of a single $PEA_2PbI_4$ flake.

In conclusion, our combined study of the angle-resolved photoluminescence and Raman spectra of single two-dimensional lead-iodide microcrystals evidenced the strong correlation of the low-frequency phonons with the band edge emission. In particular, the exciton polarization is not isotropic in the plane of the octahedra lattice, as would be expected from the electronic structure,[13] highlighting the strong influence of the exciton-phonon coupling. Here subtle differences in the organic cation, like a change in the length of the aliphatic chain, can lead to very different emission polarization and PL linewidth broadening. Our results reveal the complexity of the electron-phonon coupling and provide detailed insight on how the directionality of the phonons impacts the polarization and intensity of the band-edge emission. In particular, we find evidence that combinations of phonons with non-orthogonal directionality, along the normal [110] and diagonal [100] directions, participate in the electron-phonon coupling. One consequence is that the typically applied Bose-Einstein modeling of the thermal broadening of the PL, which is based on a single LO phonon energy, [35, 39] is not adequate to describe the thermal behavior of the 2DLPs. Furthermore, such multi-phonon coupling can alter the direction of the polarization of the emitted light. These insights on how phonon directionality impacts exciton polarization are important for designing photonic devices, in particular, where the orientation of the emission dipoles is crucial for their performance, for example in optical cavities and other resonators. Considering the complexity of the structural and vibrational properties of the two-dimensional metal halide materials, approaches using machine learning and neural-network based pattern recognition could be suitable tools to predict their optoelectronic behavior.

EXPERIMENTAL SECTION

Materials. Lead (II) iodide ($PbI_2$, 99 %), hydroiodic acid (HI, 57 %, distilled, 99.999 % trace metal basis), hypophosphorous acid ($H_3PO_2$, 50 %), phenethylamine (PEA, 99 %), undecylamine (UDA, ≥98 %), butylamine (BA, 99.5 %), ethyl acetate (≥99.5 %), toluene (≥99.7 %) were purchased from Sigma-Aldrich without any further purification.

Sample fabrication. Microcrystalline powders of $BA_2PbI_4$, $UDA_2PbI_4$ and $PEA_2PbI_4$ were synthesized by a modified antisolvent-assisted fast crystallization approach published previously by our group. [40] Briefly, 0.25 mmol $PbI_2$ was dissolved in 200 μl of HI and 108 μl of $H_3PO_2$, followed by the dilution with 2 ml ethyl acetate for $UDA_2PbI_4$ and $PEA_2PbI_4$ and 2 ml ethyl acetate and 1 ml toluene for BA$_2$PbI$_4$. Afterward, 0.6 mmol of the respective amine was directly injected into the mixture, leading to the immediate crystallization of orange microcrystals. The reaction mixture was continuously agitated for at least 3 h. Subsequently, the microcrystalline powder was separated from the solution by centrifugation at 6000 rpm, followed by the redispersion in the respective solvent system. The washing procedure was repeated twice and the purified crystals were dried under vacuum for 1 h.

Angle-resolved PL, reflectance, and Raman experiments. Raman experiments were performed under resonant conditions with wavelengths of 457 nm from an Ar$^+$ laser and non-resonant conditions with wavelengths of 633 nm from a He−Ne laser. The spectra were collected in a backscattering geometry using a Jobin–Yvon HR800 micro-Raman system equipped with a charged-coupled detector (CCD) and a 50× objective (numerical aperture (NA) of 0.55) with a long working distance. 2400 lines/ mm gratings were used yielding a spectral resolution was 0.19 cm$^{-1}$ per CCD pixel under 633 nm excitation with 2400 lines/mm. The laser plasma lines were removed by Bragg-volume-grating-based bandpass filters (BPF) from OptiGrate Corp. The samples were placed in a closed cycle cryostat (Montana Instruments) and cooled to cryogenic temperatures of about 4K. For angle-resolved Raman measurements, two vertical polarizers and one rotational half-wave plate were utilized. One polarizer was inserted in the incident path, and one analyzer with polarization parallel was allocated in the scattered path before the spectrometer. The half-wave plate was positioned in the common optical path of the incident and scattered light to simultaneously alter their polarization directions. Rotating the fast axis of the half-wave plate by an angle of $\varphi/2$ is equivalent to a rotation of the sample by an angle of $\varphi$ relative to the polarization directions of the polarizer and analyzer.[41] The laser power was kept lower than 40μW to avoid laser-induced degradation of the sample. For the temperature-dependent measurements, the sample was first cooled to 4 K and then the temperature was raised stepwise to the indicated values. The angle-resolved photoluminescence spectra were recorded with the Raman setup using a grating with 600 lines/mm and a laser at 457 nm wavelength from an $Ar^+$ laser. For PL measurements, the laser power is lower than 0.1μW. The reflectance experiments were conducted with the same setup where the laser was replaced by a white light source.

ASSOCIATED CONTENT

**Supporting Information**. The following files are available free of charge.

The Supporting_Information.pdf file contains:

- Structural properties of the 2DLPs under discussion

- Additional Raman, reflectance and PL data from $BA_2PbI_4$ flakes

- Additional Raman, reflectance and PL data from $UDA_2PbI_4$ flakes

- Angle-resolved emission, reflectance and Raman spectra of $PEA_2PbI_4$

AUTHOR INFORMATION

**Corresponding Author**

* Roman.krahne@iit.it

**Author Contributions**

The manuscript was written through contributions of all authors. All authors have given approval to the final version of the manuscript.

ACKNOWLEDGMENT

A. S. acknowledges the European Union's Horizon 2020 Research and Innovation programme under the Marie Skłodowska-Curie Funding Program (Project Together, grant agreement No.101067869). R.K acknowledges funding by the European Union under Project 101131111 – DELIGHT. P.H. and M.L. acknowledge the support from National Natural Science Foundation of China (Grant Nos. 12322401 and 12127807), Beijing Nova Program (Grant No. 20230484301), Youth Innovation Promotion Association, Chinese Academy of Sciences (No. 2023125).

REFERENCES

1. Smith, M. D.; Crace, E. J.; Jaffe, A.; Karunadasa, H. I. The Diversity of Layered Halide Perovskites, Annu. Rev. Mater. Res. 2018, 48, 111-136.
2. Tanaka, K.; Takahashi, T.; Kondo, T.; Umeda, K.; Ema, K.; Umebayashi, T.; Asai, K.; Uchida, K.; Miura, N. Electronic and Excitonic Structures of Inorganic–Organic Perovskite-Type Quantum-Well Crystal ($C_4H_9NH_3)_2PbBr_4$, Jap. J. Appl. Phys. 2005, 44, 5923.
3. Blancon, J. C.; Stier, A. V.; Tsai, H.; Nie, W.; Stoumpos, C. C.; Traoré, B.; Pedesseau, L.; Kepenekian, M.; Katsutani, F.; Noe, G. T.; Kono, J.; Tretiak, S.; Crooker, S. A.; Katan, C.; Kanatzidis, M. G.; Crochet, J. J.; Even, J.; Mohite, A. D. Scaling Law for Excitons in 2D Perovskite Quantum Wells, Nat. Commun. 2018, 9, 2254.
4. Mauck, C. M.; Tisdale, W. A. Excitons in 2D Organic–Inorganic Halide Perovskites, Trends Chem. 2019, 1, 380-393.
5. Cho, Y.; Berkelbach, T. C. Optical Properties of Layered Hybrid Organic–Inorganic Halide Perovskites: A Tight-Binding Gw-Bse Study, J. Phys. Chem. Lett. 2019, 10, 6189-6196.
6. Li, X.; Hoffman, J. M.; Kanatzidis, M. G. The 2D Halide Perovskite Rulebook: How the Spacer Influences Everything from the Structure to Optoelectronic Device Efficiency, Chem. Rev. 2021, 121, 2230-2291.
7. Smith, M. D.; Karunadasa, H. I. White-Light Emission from Layered Halide Perovskites, Acc. Chem. Res. 2018, 51, 619-627.
8. Dhanabalan, B.; Biffi, G.; Moliterni, A.; Olieric, V.; Giannini, C.; Saleh, G.; Ponet, L.; Prato, M.; Imran, M.; Manna, L.; Krahne, R.; Artyukhin, S.; Arciniegas, M. P. Engineering the Optical Emission and Robustness of Metal-Halide Layered Perovskites through Ligand Accommodation, Adv. Mater. 2021, 33, 2008004.
9. Do, T. T. H.; Granados del Águila, A.; Zhang, D.; Xing, J.; Liu, S.; Prosnikov, M. A.; Gao, W.; Chang, K.; Christianen, P. C. M.; Xiong, Q. Bright Exciton Fine-Structure in Two-Dimensional Lead Halide Perovskites, Nano Lett. 2020, 20, 5141-5148.
10. Dyksik, M.; Duim, H.; Maude, D. K.; Baranowski, M.; Loi, M. A.; Plochocka, P. Brightening of Dark Excitons in 2D Perovskites, Sci. Adv. 2021, 7, eabk0904.

11. Quarti, C.; Giorgi, G.; Katan, C.; Even, J.; Palummo, M.  Exciton Ground State Fine Structure and Excited States Landscape in Layered Halide Perovskites from Combined Bse Simulations and Symmetry Analysis, Adv. Opt. Mater. 2023, 2202801.
12. Ammirati, G.; Martelli, F.; O'Keeffe, P.; Turchini, S.; Paladini, A.; Palummo, M.; Giorgi, G.; Cinquino, M.; De Giorgi, M.; De Marco, L.; Catone, D.  Band Structure and Exciton Dynamics in Quasi-2D Dodecylammonium Halide Perovskites, Adv. Opt. Mater. 2023, 11, 2201874.
13. Giorgi, G.; Yamashita, K.; Palummo, M.  Nature of the Electronic and Optical Excitations of Ruddlesden–Popper Hybrid Organic–Inorganic Perovskites: The Role of the Many-Body Interactions, J. Phys. Chem. Lett. 2018, 9, 5891-5896.
14. Katan, C.; Mercier, N.; Even, J.  Quantum and Dielectric Confinement Effects in Lower-Dimensional Hybrid Perovskite Semiconductors, Chem. Rev. 2019, 119, 3140-3192.
15. Sercel, P. C.; Efros, A. L.  Unique Signatures of the Rashba Effect in the Magneto-Optical Properties of Two-Dimensional Semiconductors, Phys. Rev. B 2023, 107, 195436.
16. Shinde, A.; Rajput, P. K.; Makhija, U.; Tanwar, R.; Mandal, P.; Nag, A.  Emissive Dark Excitons in Monoclinic Two-Dimensional Hybrid Lead Iodide Perovskites, Nano Lett. 2023, 23, 6985-6993.
17. Folpini, G.; Cortecchia, D.; Petrozza, A.; Srimath Kandada, A. R.  The Role of a Dark Exciton Reservoir in the Luminescence Efficiency of Two-Dimensional Tin Iodide Perovskites, J. Mater. Chem. C 2020, 8, 10889-10896.
18. Posmyk, K.; Dyksik, M.; Surrente, A.; Maude, D. K.; Zawadzka, N.; Babiński, A.; Molas, M. R.; Paritmongkol, W.; Mączka, M.; Tisdale, W. A.; Plochocka, P.; Baranowski, M.  Exciton Fine Structure in 2d Perovskites: The Out-of-Plane Excitonic State, Adv. Opt. Mater. 2023, 2300877.
19. Straus, D. B.; Kagan, C. R.  Photophysics of Two-Dimensional Semiconducting Organic–Inorganic Metal-Halide Perovskites, Annu. Rev. Phys. Chem. 2022, 73, 403-428.
20. Straus, D. B.; Hurtado Parra, S.; Iotov, N.; Zhao, Q.; Gau, M. R.; Carroll, P. J.; Kikkawa, J. M.; Kagan, C. R.  Tailoring Hot Exciton Dynamics in 2D Hybrid Perovskites through Cation Modification, ACS Nano 2020, 14, 3621-3629.
21. Straus, D. B.; Hurtado Parra, S.; Iotov, N.; Gebhardt, J.; Rappe, A. M.; Subotnik, J. E.; Kikkawa, J. M.; Kagan, C. R.  Direct Observation of Electron–Phonon Coupling and Slow Vibrational Relaxation in Organic–Inorganic Hybrid Perovskites, J. Am. Chem. Soc. 2016, 138, 13798-13801.
22. Thouin, F.; Valverde-Chávez, D. A.; Quarti, C.; Cortecchia, D.; Bargigia, I.; Beljonne, D.; Petrozza, A.; Silva, C.; Srimath Kandada, A. R.  Phonon Coherences Reveal the Polaronic Character of Excitons in Two-Dimensional Lead Halide Perovskites, Nat. Mater. 2019, 18, 349-356.
23. Thouin, F.; Srimath Kandada, A. R.; Valverde-Chávez, D. A.; Cortecchia, D.; Bargigia, I.; Petrozza, A.; Yang, X.; Bittner, E. R.; Silva, C.  Electron–Phonon Couplings Inherent in Polarons Drive Exciton Dynamics in Two-Dimensional Metal-Halide Perovskites, Chem. Mater. 2019, 31, 7085-7091.
24. Srimath Kandada, A. R.; Silva, C.  Exciton Polarons in Two-Dimensional Hybrid Metal-Halide Perovskites, J. Phys. Chem. Lett. 2020, 11, 3173-3184.
25. Dyksik, M.; Beret, D.; Baranowski, M.; Duim, H.; Moyano, S.; Posmyk, K.; Mlayah, A.; Adjokatse, S.; Maude, D. K.; Loi, M. A.; Puech, P.; Plochocka, P.  Polaron Vibronic

Progression Shapes the Optical Response of 2D Perovskites, Advanced Science 2023, 2305182.
26. Mauck, C. M.; France-Lanord, A.; Hernandez Oendra, A. C.; Dahod, N. S.; Grossman, J. C.; Tisdale, W. A. Inorganic Cage Motion Dominates Excited-State Dynamics in 2d-Layered Perovskites $(C_xH_{2x+1}NH_3)_2PbI_4$ (X = 4–9), J. Phys. Chem. C 2019, 123, 27904-27916.
27. Posmyk, K.; Dyksik, M.; Surrente, A.; Zalewska, K.; Śmiertka, M.; Cybula, E.; Paritmongkol, W.; Tisdale, W. A.; Plochocka, P.; Baranowski, M., Fine Structure Splitting of Phonon-Assisted Excitonic Transition in $(PEA)_2PbI_4$ Two-Dimensional Perovskites. Nanomaterials, 2023; 13, 1119.
28. Posmyk, K.; Zawadzka, N.; Dyksik, M.; Surrente, A.; Maude, D. K.; Kazimierczuk, T.; Babiński, A.; Molas, M. R.; Paritmongkol, W.; Mączka, M.; Tisdale, W. A.; Płochocka, P.; Baranowski, M. Quantification of Exciton Fine Structure Splitting in a Two-Dimensional Perovskite Compound, J. Phys. Chem. Lett. 2022, 13, 4463-4469.
29. Pérez-Osorio, M. A.; Lin, Q.; Phillips, R. T.; Milot, R. L.; Herz, L. M.; Johnston, M. B.; Giustino, F. Raman Spectrum of the Organic–Inorganic Halide Perovskite $CH_3NH_3PbI_3$ from First Principles and High-Resolution Low-Temperature Raman Measurements, J. Phys. Chem. C 2018, 122, 21703-21717.
30. Pérez-Osorio, M. A.; Milot, R. L.; Filip, M. R.; Patel, J. B.; Herz, L. M.; Johnston, M. B.; Giustino, F. Vibrational Properties of the Organic–Inorganic Halide Perovskite $CH_3NH_3PbI_3$ from Theory and Experiment: Factor Group Analysis, First-Principles Calculations, and Low-Temperature Infrared Spectra, J. Phys. Chem. C 2015, 119, 25703-25718.
31. Dhanabalan, B.; Leng, Y.-C.; Biffi, G.; Lin, M.-L.; Tan, P.-H.; Infante, I.; Manna, L.; Arciniegas, M. P.; Krahne, R. Directional Anisotropy of the Vibrational Modes in 2d-Layered Perovskites, ACS Nano 2020, 14, 4689-4697.
32. Martín-García, B.; Spirito, D.; Lin, M.-L.; Leng, Y.-C.; Artyukhin, S.; Tan, P.-H.; Krahne, R. Low-Frequency Phonon Modes in Layered Silver-Bismuth Double Perovskites: Symmetry, Polarity, and Relation to Phase Transitions Adv. Opt. Mater. 2022, 10, 2270056.
33. Lin, M.-L.; Dhanabalan, B.; Biffi, G.; Leng, Y.-C.; Kutkan, S.; Arciniegas, M. P.; Tan, P.-H.; Krahne, R. Correlating Symmetries of Low-Frequency Vibrations and Self-Trapped Excitons in Layered Perovskites for Light Emission with Different Colors, Small 2022, 18, 2106759.
34. Lemmerer, A.; Billing, D. G. Synthesis, Characterization and Phase Transitions of the Inorganic–Organic Layered Perovskite-Type Hybrids $[(CNH_{2n+1}NH_3)_2PbI_4]$, N = 7, 8, 9 and 10, Dalton Trans. 2012, 41, 1146-1157.
35. Wright, A. D.; Verdi, C.; Milot, R. L.; Eperon, G. E.; Pérez-Osorio, M. A.; Snaith, H. J.; Giustino, F.; Johnston, M. B.; Herz, L. M. Electron-Phonon Coupling in Hybrid Lead Halide Perovskites, Nat. Commun. 2016, 7, 1-9.
36. Anantharaman, S. B.; Lynch, J.; Stevens, C. E.; Munley, C.; Li, C.; Hou, J.; Zhang, H.; Torma, A.; Darlington, T.; Coen, F.; Li, K.; Majumdar, A.; Schuck, P. J.; Mohite, A.; Harutyunyan, H.; Hendrickson, J. R.; Jariwala, D. Dynamics of Self-Hybridized Exciton–Polaritons in 2d Halide Perovskites, Light: Science & Applications 2024, 13, 1.
37. Gan, L.; Li, J.; Fang, Z.; He, H.; Ye, Z. Effects of Organic Cation Length on Exciton Recombination in Two-Dimensional Layered Lead Iodide Hybrid Perovskite Crystals, J. Phys. Chem. Lett. 2017, 8, 5177-5183.

38. Kahmann, S.; Duim, H.; Fang, H.-H.; Dyksik, M.; Adjokatse, S.; Rivera Medina, M.; Pitaro, M.; Plochocka, P.; Loi, M. A.  Photophysics of Two-Dimensional Perovskites—Learning from Metal Halide Substitution, Adv. Funct. Mater. 2021, 31, 2103778.
39. Kutkan, S.; Dhanabalan, B.; Lin, M.-L.; Tan, P.-H.; Schleusener, A.; Arciniegas, M. P.; Krahne, R.  Impact of the Organic Cation on the Band-Edge Emission of Two-Dimensional Lead–Bromide Perovskites, Nanoscale 2023, 15, 12880-12888.
40. Dhanabalan, B.; Castelli, A.; Palei, M.; Spirito, D.; Manna, L.; Krahne, R.; Arciniegas, M.  Simple Fabrication of Layered Halide Perovskite Platelets and Enhanced Photoluminescence from Mechanically Exfoliated Flakes, Nanoscale 2019, 11, 8334-8342.
41. Liu, X.-L.; Zhang, X.; Lin, M.-L.; Tan, P.-H.  Different Angle-Resolved Polarization Configurations of Raman Spectroscopy: A Case on the Basal and Edge Plane of Two-Dimensional Materials, Chin. Phys. B 2017, 26, 067802.

38. Kahmann, S.; Duim, H.; Fang, H.-H.; Dyksik, M.; Adjokatse, S.; Rivera Medina, M.; Pitaro, M.; Plochocka, P.; Loi, M. A.  Photophysics of Two-Dimensional Perovskites—Learning from Metal Halide Substitution, Adv. Funct. Mater. 2021, 31, 2103778.
39. Kutkan, S.; Dhanabalan, B.; Lin, M.-L.; Tan, P.-H.; Schleusener, A.; Arciniegas, M. P.; Krahne, R.  Impact of the Organic Cation on the Band-Edge Emission of Two-Dimensional Lead–Bromide Perovskites, Nanoscale 2023, 15, 12880-12888.
40. Dhanabalan, B.; Castelli, A.; Palei, M.; Spirito, D.; Manna, L.; Krahne, R.; Arciniegas, M.  Simple Fabrication of Layered Halide Perovskite Platelets and Enhanced Photoluminescence from Mechanically Exfoliated Flakes, Nanoscale 2019, 11, 8334-8342.
41. Liu, X.-L.; Zhang, X.; Lin, M.-L.; Tan, P.-H.  Different Angle-Resolved Polarization Configurations of Raman Spectroscopy: A Case on the Basal and Edge Plane of Two-Dimensional Materials, Chin. Phys. B 2017, 26, 067802.

TOC image

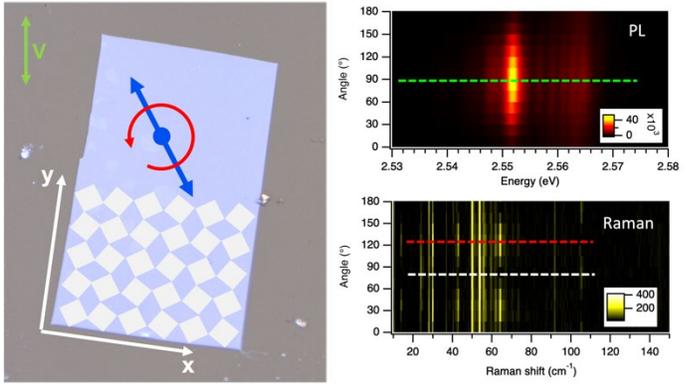